\documentstyle[longtable]{aipproc}
\input{epsf}
\begin{document}

\title{Asymmetry studies in $\Lambda^0/\bar{\Lambda}^0$, $\Xi^-/\Xi^+$ 
and $\Omega^-/\Omega^+$ production}

\author{J.C. Anjos, J. Magnin~\thanks{supported by FAPERJ, 
Funda\c{c}\~ao de Amparo \`a Pesquisa do Estado de Rio de Janeiro.}, 
F.R.A. Sim\~ao and J. Solano}
\address{Centro Brasileiro de Pesquisas F\'{\i}sicas - CBPF\\
Rua Dr. Xavier Sigaud 150, CEP 22290-180, Rio de Janeiro, Brazil.}
\maketitle


\begin{abstract}
We present a study on hyperon/anti-hyperon production asymmetries in 
the framework of the recombination model. The production asymmetries 
for $\Lambda^0/\bar{\Lambda}^0$, $\Xi^-/\Xi^+$ and $\Omega^-/\Omega^+$ 
are studied as a function of $x_F$. Predictions of the model are 
compared to preliminary data on hyperon/anti-hyperon production asymmetries 
in $500$ GeV/$c$ $\pi^-p$ interactions from the Fermilab E791 experiment. 
The model predicts a growing asymmetry with the number of valence 
quarks shared by the target and the produced hyperons in the $x_F < 0$ 
region. In the positive $x_F$ region, the model predicts constant 
asymmetries for $\Lambda^0/\bar{\Lambda}^0$ and $\Omega^-/\Omega^+$ 
production and a growing asymmetry with $x_F$ for $\Xi^-/\Xi^+$. 
We found a qualitatively good agreement between the model predictions and 
data, showing that recombination is a competitive mechanism 
in the hadronization process.
\end{abstract}

\section{Introduction}

In hadron interactions, the leading particle effect manifests as an 
enhancement in the production 
rate of particles which share valence quarks with the initial hadrons. 
As a consequence of the leading particle effect, 
strong asymmetries are expected in the $x_F\; (=2p_L/\sqrt{s})$ inclusive 
differential cross sections for particles and anti-particles when the content 
of valence quarks shared by the produced particle and anti-particle with the 
initial hadrons is different. 

This effect has been extensively studied, from both the 
experimental~\cite{charm-ex} and theoretical~\cite{charm-th1,charm-th2} 
points of view in charm hadron production. 

In the production of strange baryons, the same type of leading effects 
are expected. Indeed, there is some evidence of asymmetries in 
$\Lambda^0 /\overline{\Lambda}^0$ production in $\pi^-$Cu interactions 
at 230 GeV/$c$~\cite{accmor1} and in $\Lambda^0 /\bar{\Lambda}^0$ and 
$\Xi^-/\Xi^+$ production in 250 GeV/$c$ $\pi^-p$ interactions~\cite{bogert}. 
Some additional evidence for $\Lambda^0 /\bar{\Lambda}^0$ asymmetry 
can be found in Ref.~\cite{add-evid}, but, in general, hyperon 
production asymmetries in $\pi^-p$ interactions were not systematically 
studied until recently, when the Fermilab E791 experiment presented 
preliminary results on hyperon production in $\pi^-p$ interactions at 
$500$ GeV/$c$~\cite{ranp97}. The E791 experiment 
has measured particle/anti-particle production asymmetries in both 
the small $x_F<0$ and $x_F>0$ regions for the $\Lambda^0/\bar{\Lambda}^0$, 
$\Xi^-/\Xi^+$ and $\Omega^-/\Omega^+$ hyperons.

The results obtained by the E791 experiment show a large asymmetry 
in the $x_F<0$ region for $\Lambda^0/\bar{\Lambda}^0$ production and 
a lower asymmetry, in the same region, for $\Xi^-/\Xi^+$ production. 
In the $x_F>0$ region, an approximately constant asymmetry with $x_F$ 
is observed for $\Xi^-/\Xi^+$ and $\Lambda^0/\bar{\Lambda}^0$ 
hyperons. The asymmetry measured for $\Omega^-/\Omega^+$ production 
is approximately constant in the whole $-0.12 \leq x_F \leq 0.12$ region.

These results in the $x_F<0$ region are consistent with the fact 
that $\Lambda^0$ hyperons share a $ud$ diquark whereas the $\Xi^-$s 
share a $d$ quark with the target particles (protons and neutrons), so 
a lower asymmetry is expected for the later since the $\Lambda^0$ is a double 
leading whereas the $\Xi^-$ is a leading particle. In the $x_F>0$ region, 
the $\Xi^-$ share a $d$ quark with the initial $\pi^-$, being 
a leading particle, while $\Xi^+$ shares none. Then a growing asymmetry with 
$x_F$ is expected in $\Xi^-/\Xi^+$ production. The $\Lambda^0$ and $\bar{\Lambda}^0$ each have one valence quark in common with the beam 
$\pi^-$ and, therefore, have equal enhancement i.e., no asymmetry is 
expected from this effect. 

The $\Omega^-/\Omega^+$ hyperons are both non-leading 
in all the $x_F$ regions studied and, consequently, no asymmetry is 
expected in $\Omega^-/\Omega^+$ production in $\pi^-p$ interactions.

Due to the smallness of the strange quark mass and the $p_T^2$ values
involved, hyperon production can not be accounted for in the usual 
framework of perturbative QCD.

The recombination scheme, initially introduced by Das and 
Hwa~\cite{das-hwa}, appears to be a possible framework to deal with the 
non-perturbative QCD aspects involved in hadron and, in particular, in 
hyperon production. Indeed, this type of model has been used succesfuly 
to describe charmed particle/anti-particle asymmetries~\cite{charm-th2,hwa95} 
in hadroproduction.

In this work, we compare the predictions of a simple version of the 
recombination model on hyperon/anti-hyperon production asymmetries 
with the preliminary results of the E791 experiment. The good qualitative 
agreement found between model predictions and experimental data even 
at small values of $x_F$ shows that it might be interesting to make efforts 
to improve the outcome of the model.

\section{Hyperon production by recombination in $\pi^-p$ interactions}

The recombination model was introduced long time ago by Das 
and Hwa~\cite{das-hwa} to describe meson production in 
hadron-hadron collisions. A simple extension of the model was made by 
Ranft~\cite{ranft} to calculate baryon production. From those first 
attempts up to now, several modifications have been introduced 
trying to improve the outcome of the model~\cite{improvements}. 
Nevertheless, the recombination model remains a simple approach to deal 
with some non-perturbative aspects of QCD involved in hadron-hadron 
interactions.

The basic idea behind recombination is that the produced hadrons are 
formed from the debris of the fragmented beam (in the forward region) or target 
(in the backward direction) particles in such a way that partons initially 
in the incoming particles {\em recombine} into the final hadrons. 
All that is needed to deal with the problem is to know the distribution 
of partons in the initial particles, which are measured in Deep Inelastic 
Scattering experiments, and the so-called recombination function which will 
take into account all aspects involved in the recombination of partons into 
a hadron. Of course, the recombination function has a phenomenological 
origin, since no calculation from first principles is yet possible to 
obtain it. 

For a generic hyperon $H$, the $x_F$ inclusive distribution in recombination 
is given by
\begin{equation} 
\frac{2 E}{\sigma^{rec}\sqrt {s}}\frac{d\sigma^{rec}
}{d\left|x_F\right|}=\int_0^1\frac{dx_1}{x_1}\frac{dx_2}{x_2}\frac{dx_3
}{x_3}F_3^{H}
\left( x_1,x_2,x_3\right) R_3\left( x_1,x_2,x_3,x_F\right) \; ,
\label{rec}
\end{equation}
where $\sqrt{s}$ is the center of mass energy in the $\pi^-p\rightarrow 
H+X$ reaction, $E$ is the energy of the outgoing hyperon and $\sigma^{rec}$ 
is a normalization constant. 
In eq.~(\ref{rec}), $F_3^{H}\left( x_1,x_2,x_3\right)$ is the 
three-quark distribution, which contains the distribution of valence 
quarks in the final particle inside the beam or target hadrons, 
$R_3\left( x_1,x_2,x_3,x_F\right)$ is the recombination function and 
$x_i;\; i=1,2,3$ is the momentum fraction of the $i^{th}$ quark with respect 
to the initial particle. 

Following the approach of Ref.~\cite{ranft}, the three quark 
distribution function is assumed to be of the form
\begin{equation}
F_3^{H} \left( x_1,x_2,x_3 \right) = 
\beta g(x_1,x_2,x_3) \left(1-x_1-x_2-x_3\right)^{\gamma} \; , 
\label{3quark}
\end{equation}
where
\begin{equation}
g(x_1,x_2,x_3) = F_{q_1}\left(x_1\right)F_{q_2}\left(x_2\right)
F_{q_3}\left(x_3\right)
\label{3quark2}
\end{equation}
contains the single quark distribution, $F_{q_i} = x_i q_i(x_i)$, 
of the $q_i$ valence quark of the final hyperon in the initial particle. 
Note that $F_{q_i}$ includes valence as well as sea quark contributions 
from the initial hadron. The coefficients $\beta$ and $\gamma$ are fixed 
using the consistency condition
\begin{eqnarray}
F_{q}\left(x_i\right) & = & \int_0^{1-x_i}dx_j \int_0^{1-x_i-x_j}
dx_k \:F_3^{H} 
\left( x_1,x_2,x_3 \right)\ \nonumber \\
                      &   &i,j,k = 1,2,3 
\label{consist}
\end{eqnarray}
which must be valid for the valence quarks in the initial particle.

For the recombination function we use~\cite{plb96}
\begin{equation}
R_3\left( x_1,x_2,x_3\right) =\alpha \frac{(x_1x_2)^{n_1}x_3^{n_2}}
{x_F^{n_1+n_2-1}}
\delta \left(x_1+x_2+x_3-x_F\right)   
\label{eq7}
\end{equation}
allowing in this way a different weight for the heavier $s$ ($\bar{s}$) 
quark than for the light $u$ ($\bar{u}$) and $d$ ($\bar{d}$) quarks. 

The constant $\alpha$ in eq.~(\ref{eq7}) is fixed by the condition 
\cite{nosotros} 
\begin{equation}
\frac{1}{\sigma^{rec}} 
\int_0^1 dx_F \frac{d\sigma^{rec}}{dx_F} = 1 \;,
\label{eq7b}
\end{equation}
then $\sigma^{rec}$ is the recombination cross section of the hyperon $H$ 
in $\pi^-p\rightarrow H+X$ in the forward ($x_F>0$) or backward ($x_F<0$) 
region. $\sigma^{rec}$ may be fixed from experimental data.

The asymmetry as a function of $x_F$ is defined  by
\begin{equation}
A(x_F) = \frac{d\sigma_{H}/d\left|x_F\right| - 
d\sigma_{\bar{H}}/d\left|x_F\right|}{d\sigma_{H}/d\left|x_F\right| + 
d\sigma_{\bar{H}}/d\left|x_F\right|}
\label{eq0}
\end{equation}
where $H$ is the Hyperon and $\bar{H}$ is the anti-Hyperon. 

Replacing eq.~(\ref{rec}), with eqs.~(\ref{3quark}) to (\ref{eq7b}), 
into eq.~(\ref{eq0}) for the hyperons and anti-hyperons we obtain
\begin{eqnarray}
A(x_F) & = & \frac{
\int_0^{|x_F|}dx_1 {\int_0^{|x_F|-x_1}dx_2
\left[ g_{H}\left(x_1,x_2,x_3\right) - \sigma 
g_{\bar{H}}\left(x_1,x_2,x_3\right)\right]}}
{\int_0^{|x_F|}dx_1 {\int_0^{|x_F|-x_1}dx_2
\left[ g_{H}\left(x_1,x_2,x_3\right) + 
\sigma g_{\bar{H}}\left(x_1,x_2,x_3\right)\right]}} \nonumber \\
x_3 & = & |x_F|-x_1-x_2
\label{quelo}
\end{eqnarray}
with $\sigma = \sigma^{\bar{H}}/\sigma^{H}$ the relative normalization 
between the hyperon and anti-hyperon distributions. The delta 
function of eq.~(\ref{eq7}) has been used to do one of the integrals in 
eq.~(\ref{quelo}).

As in Ref.~\cite{plb96}, we have used $n_1=1$, $n_2=3/2$ and 
$\gamma = -0.3$ in our calculations. 

\begin{figure}[t]
\begin{minipage}{2.7in}
\begin{center}
\epsfxsize=2.7in \epsfbox{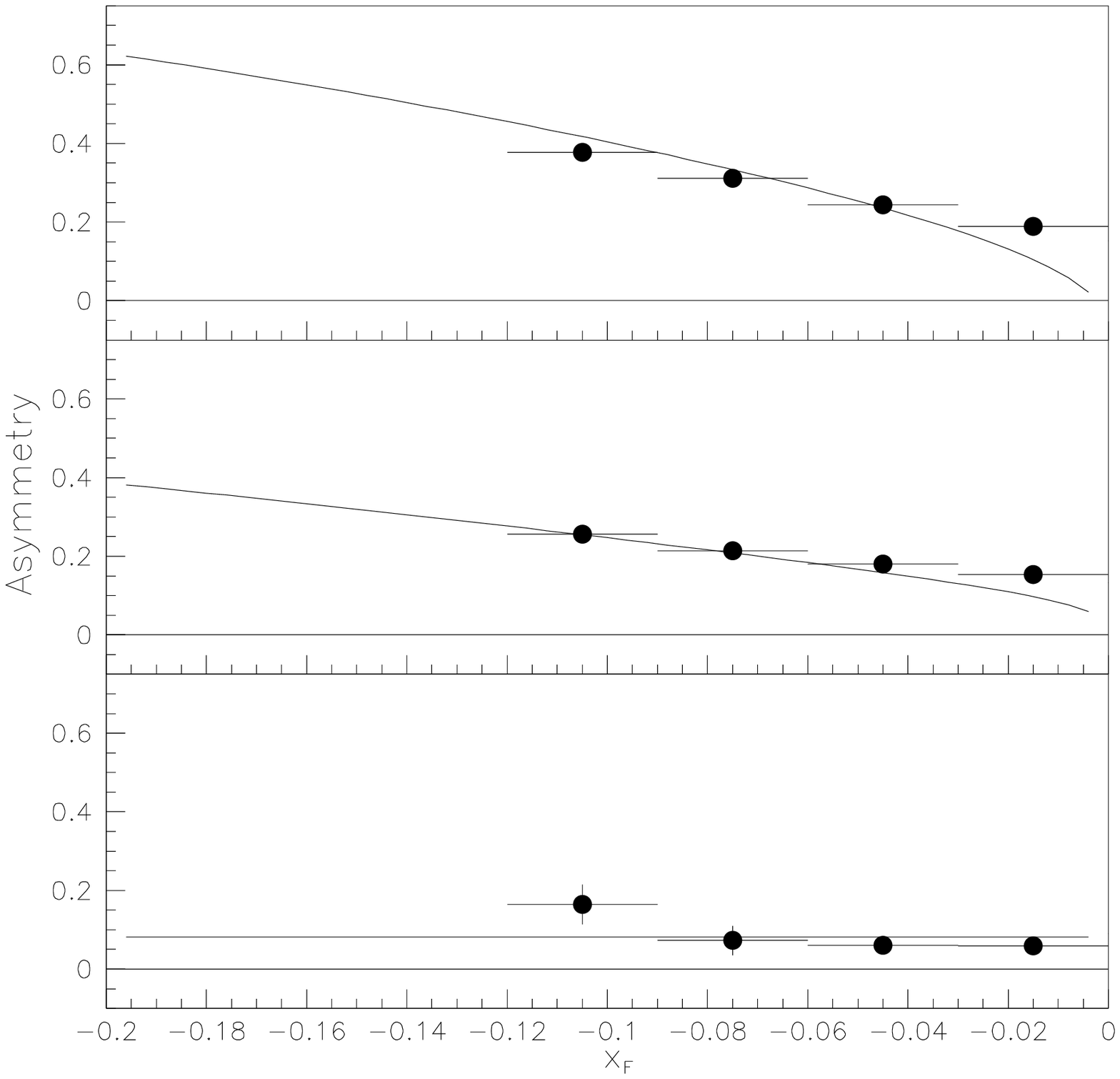}
\caption{$\Lambda^0/\bar{\Lambda}^0$ (upper), $\Xi^-/\Xi^+$ (middle) 
and $\Omega^-/\Omega^+$ (lower) asymmetries as a function of $x_F$ in the 
negative $x_F$ region. Full line is the prediction from recombination model.  
Black dots are the preliminary E791 results taken from Ref.~[7].}
\label{fig1}
\end{center}
\end{minipage}
\ \hspace{15pt}\
\begin{minipage}{2.7in}
\begin{center}
\epsfxsize=2.7in \epsfbox{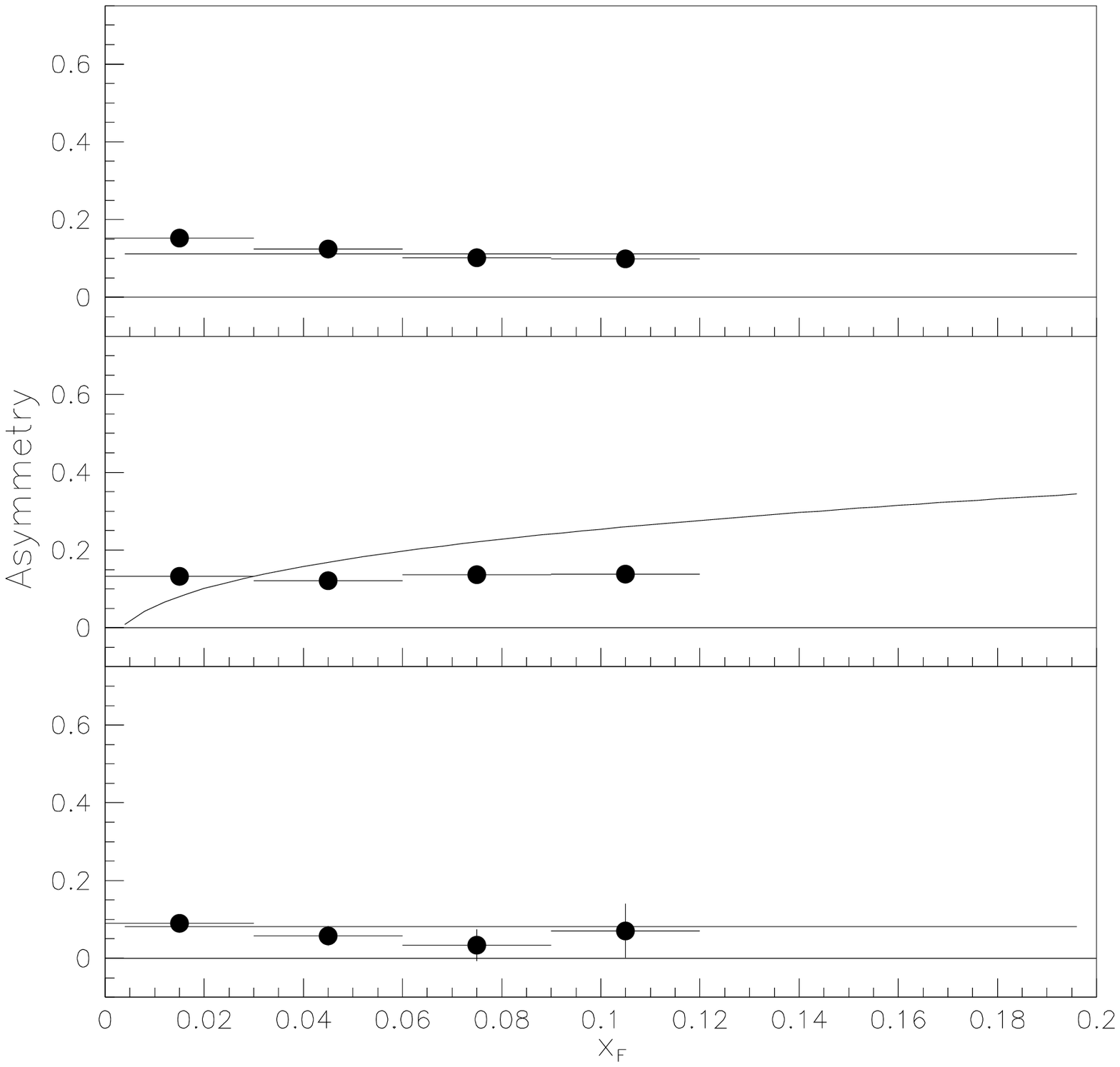}
\caption{$\Lambda^0/\bar{\Lambda}^0$ (upper), $\Xi^-/\Xi^+$ (middle) 
and $\Omega^-/\Omega^+$ (lower) asymmetries as a function of $x_F$ in the 
positive $x_F$ region. Full line is the prediction from recombination model.  
Black dots are the preliminary E791 results taken from Ref.~[7].}
\label{fig2}
\end{center}
\end{minipage}
\end{figure}

%
\begin{table}[t]
\caption{$g_{H}(x_1,x_2,x_3)$ and $g_{\bar{H}}(x_1,x_2,x_3)$ used 
in the calculation of asymmetries. $q^n$ ($\bar{q}^n$) is the quark 
(anti-quark) distribution in nucleons, $q^{\pi}$ ($\bar{q}^{\pi}$) is the 
quark (anti-quark) distribution in the $\pi^-$. The individual parton 
distributions were taken from Ref.~[12].}
\begin{tabular}{|c||c|c||c|c||} \hline\hline
  & \multicolumn{2}{|c|}{$ x_F < 0. $} & \multicolumn{2}{|c|}
{$ 0. < x_F $} \\ \hline
  & $g_{H}(x_1,x_2,x_3)$ & $g_{\bar{H}}(x_1,x_2,x_3)$ & 
$g_{H}(x_1,x_2,x_3)$ & $g_{\bar{H}}(x_1,x_2,x_3)$ \\ \hline
$\Lambda^0/\bar{\Lambda}^0$ & 
$u^n(x_1)d^n(x_2)s^n(x_3)$ & 
$\bar{u}^n(x_1)\bar{d}^n(x_2)\bar{s}^n(x_3)$ &
$u^{\pi}(x_1)d^{\pi}(x_2)s^{\pi}(x_3)$ &
$\bar{u}^{\pi}(x_1)\bar{d}^{\pi}(x_2)\bar{s}^{\pi}(x_3)$ \\ \hline
$\Xi^-/\Xi^+$ & 
$d^n(x_1)s^n(x_2)s^n(x_3)$ & 
$\bar{d}^n(x_1)\bar{s}^n(x_2)\bar{s}^n(x_3)$ &
$d^{\pi}(x_1)s^{\pi}(x_2)s^{\pi}(x_3)$ &
$\bar{d}^{\pi}(x_1)\bar{s}^{\pi}(x_2)\bar{s}^{\pi}(x_3)$ \\ \hline
$\Omega^-/\Omega^+$ & 
$s^n(x_1)s^n(x_2)s^n(x_3)$ & 
$\bar{s}^n(x_1)\bar{s}^n(x_2)\bar{s}^n(x_3)$ &
$s^{\pi}(x_1)s^{\pi}(x_2)s^{\pi}(x_3)$ &
$\bar{s}^{\pi}(x_1)\bar{s}^{\pi}(x_2)\bar{s}^{\pi}(x_3)$ \\ \hline
\end{tabular}
\label{table1}
\end{table}
%
In Figs.~\ref{fig1} and~\ref{fig2} we show the predictions of the 
recombination model compared to the E791 measurements. 
In order to obtain the theoretical curves shown in the figures, the 
asymmetry as given by eq.~(\ref{quelo}) has been calculated for 
each $x_F$ region independently. In Table~\ref{table1}, the $g_{H}$ and 
$g_{\bar{H}}$ distributions used in each region and for each one of the 
produced hyperons and anti-hyperons are displayed. The relative 
normalization between 
the particle and anti-particle $x_F$ distributions, $\sigma$, has been 
chosen to fit the experimental data (See Table~\ref{table2}).

For $\Omega^-/\Omega^+$ production, since both hyperon and anti-hyperon are 
non-leading in the $x_F<0$ as well as in the $x_F>0$ regions, the 
recombination model predicts a constant asymmetry arising only from 
the difference between the global normalization of the particle and 
anti-particle $x_F$ distributions.
\begin{table}[b]
\caption{Relative normalization, $\sigma$, between hyperon and 
anti-hyperon cross sections}
\begin{tabular}{|c|c|c|} \hline\hline
  & $ x_F < 0. $ & $ 0. < x_F $ \\ \hline
$\Lambda^0/\bar{\Lambda}^0$ & $3.5$ &  $0.8$ \\ \hline
$\Xi^-/\Xi^+$ & $1.9$ & $2.2$ \\ \hline
$\Omega^-/\Omega^+$ & $0.85$ & $0.85$ \\ \hline
\end{tabular}
\label{table2}
\end{table}

\section{Conclusions}

As can be seen in Figs.~\ref{fig1} and \ref{fig2}, the predictions of 
the recombination model agree qualitatively well with the experimental data.

In the negative $x_F$ region, the growth of the asymmetry predicted by 
the model with the number of valence quarks shared by the leading hyperons 
and the target particles is in a remarkable agreement with the E791 data. 
In this region, the asymmetry as a function of $x_F$ is also qualitatively 
well described by the model. The agreement between data and model predictions 
is better as $|x_F|$ rises. This is possibly due to the fact that, 
at very low values of $|x_F|$, other mechanisms than recombination are 
competitive in the hadronization.

In the positive $x_F$ region, although the leading effect in $\Xi^-/\Xi^+$ 
production is qualitatively accounted for by the model, in general, 
the agreement between data and recombination model predictions is poorer 
than in the negative $x_F$ region. Note, however, that the quark 
distributions in pions are not as well known as in nucleons, being one 
of the possible causes of the discrepancies between model prediction 
and data in this region.

The asymmetry predicted for $\Omega^-/\Omega^+$ production is constant 
over all the $x_F$ region under study. This is consistent with the fact 
that the $\Omega^-$ and the $\Omega^+$ are both non-leading particles over 
the whole $x_F$ region from $-1$ to $1$.

In conclusion, the recombination model, although simple, is 
able to reproduce qualitatively the behaviour of the experimental data 
on hyperon/anti-hyperon asymmetries, so it might be possibly of interest 
to make efforts in order to improve the outcome of the model. 
 
For a meaningful quantitative comparison between data and model 
predictions, however, the individual inclusive $x_F$ distributions for 
leading and non-leading particles must be taken into account.

\section*{Acknowledgements}
Two of us, J.C.A. and F.R.A.S. would like to thank the organizing 
committee and the Centro Latino Americano de Fisica, CLAF, for 
financial support to attend the SILAFAE II. The authors also would like  
to acknowledge J.A. Appel, B. Meadows and D.A. Sanders for useful 
comments.


\end{document}